\def\year{2022}\relax
\documentclass[letterpaper]{article} 
\usepackage{aaai22}  
\usepackage{times}  
\usepackage{helvet}  
\usepackage{courier}  
\usepackage[hyphens]{url}  
\usepackage{graphicx} 
\usepackage{natbib}  
\usepackage{caption} 
\DeclareCaptionStyle{ruled}{labelfont=normalfont,labelsep=colon,strut=off} 
\usepackage{algorithm}
\usepackage{algorithmic}
\usepackage{tabularx}
\usepackage{booktabs}
\usepackage{multirow}
\usepackage{graphicx}
\usepackage[normalem]{ulem}
\useunder{\uline}{\ul}{}
\usepackage{amsmath} 
\usepackage{xcolor}

\usepackage{tcolorbox}
\usepackage{newfloat}
\usepackage{listings}
\floatstyle{ruled}
\newfloat{listing}{tb}{lst}{}
\floatname{listing}{Listing}

\nocopyright
\title{Sympathy over Polarization: A Computational Discourse Analysis of Social Media Posts about the July 2024 Trump Assassination Attempt}
\author{
    Qingcheng Zeng\textsuperscript{\rm 1}\equalcontrib,
    Guanhong Liu\textsuperscript{\rm 2}\equalcontrib,
    Zhaoqian Xue\textsuperscript{\rm 3},
    Diego Ford\textsuperscript{\rm 4},
    Rob Voigt\textsuperscript{\rm 1},
    Loni Hagen\textsuperscript{\rm 4},
    Lingyao Li\textsuperscript{\rm 4}
    \thanks{Corresponding authors: qingchengzeng2027@u.northwestern.edu, lingyaol@usf.edu}
}

\affiliations{
    \textsuperscript{\rm1}Northwestern University,
    \textsuperscript{\rm2}University of Chicago,
    \textsuperscript{\rm3}Georgetown University,
    \textsuperscript{\rm4}University of South Florida\\

    qingchengzeng2027@u.northwestern.edu, guanhongliu@uchicago.edu, zx136@georgetown.edu, diegoford@usf.edu, robvoigt@northwestern.edu, lonihagen@usf.edu, lingyaol@usf.edu
}

\begin{document}

\maketitle

\begin{abstract}
On July 13, 2024, at the Trump rally in Pennsylvania, someone attempted to assassinate Republican Presidential Candidate Donald Trump. This attempt sparked a large-scale discussion on social media. We collected posts from X (formerly known as Twitter) one week before and after the assassination attempt and aimed to model the short-term effects of such a ``shock'' on public opinions and discussion topics. Specifically, our study addresses three key questions: first, we investigate how public sentiment toward Donald Trump shifts over time and across regions (RQ1) and examine whether the assassination attempt itself significantly affects public attitudes, independent of the existing political alignments (RQ2). Finally, we explore the major themes in online conversations before and after the crisis, illustrating how discussion topics evolved in response to this politically charged event (RQ3). By integrating large language model-based sentiment analysis, difference-in-differences modeling, and topic modeling techniques, we find that following the attempt the public response was broadly sympathetic to Trump rather than polarizing, despite baseline ideological and regional disparities.
\end{abstract}

\section{Introduction} 
Shock events, whether health emergencies or political upheavals, often trigger dramatic changes in public opinion toward key figures and even political choices \cite{johansson2021rally, mackintosh2014crises}. Among these, assassination attempts targeting public figures stand out as one of the most extreme forms of ``shock events,'' often attracting intense media attention and provoking strong emotional reactions \cite{pillemer2000momentous, atkeson2012catastrophic}. The attempted assassination of Donald Trump in Pennsylvania in July 2024 exemplifies how such events can affect partisan attitudes. Using national survey data, \citet{holliday2024trump} found that perhaps surprisingly this event did not inflame partisan tensions; Democrats' attitudes were largely unchanged by the event, while Republicans expressed reduced support for partisan violence following the incident and did not significantly decrease in their attitudes towards the out-party. 

While these findings speak to changes in attitudes, it is still unknown whether changes in the public discourse surrounding crisis events follow similar patterns. In this work we extend \citet{holliday2024trump} into the realm of public discourse by conducting a thorough case study on social media posts following the attempted assassination of Donald Trump in July 2024.  
Existing literature suggests two contrasting possibilities for the impact of such events on public sentiment, which we characterize as a distinction between sympathy versus polarization. Some research suggests sympathy may predominate, given documented temporary surges in public approval for political leaders following traumatic events like assassination attempts. This response is typically attributed to public sympathy for the leader’s personal ordeal rather than evaluations of their policy performance or partisan alignment \cite{ostrom1985promise}, as in the case of Ronald Reagan’s surge in popularity following the assassination attempt in 1981 \cite{gilbert2013attempted}. Polarization, by contrast, highlights the potential for such events to deepen political and ideological divides \cite{whitlinger2019political}, as observed in the aftermath of Martin Luther King Jr.’s assassination, which intensified racial tensions and radicalized both supporters and opponents of civil rights \cite{putnam2007e, sokol2018heavens, whitlinger2019political}. The sympathy hypothesis suggests a unifying response with broad-based support for Trump across political groups, while the polarization hypothesis predicts a divisive outcome, deepening partisan identities and ideological rifts. 

While survey data can provide structured and representative insights into such public opinion change,
its collection can be resource-intensive and time-consuming \cite{groves2009survey}. By contrast, social media data provides real-time, large-scale reflections of public sentiment, capturing dynamic reactions to political shocks \cite{gonzalez2014assessing, pandarachalil2015twitter}. The availability of metadata such as timestamps and geolocations enables temporal and spatial analysis of sentiment trends, offering insights into regional and demographic variation \cite{bollen2010twitter, pang2008opinion}. In the context of the United States’ deep political polarization, social media has been demonstrated a valuable channel to understand the geographic and ideological differences in responses to political shock events \cite{bennett2021mapping, rueda2023disaggregated}.


To investigate the impact of this political shock on public sentiment, we employ advanced natural language processing (NLP) and statistical techniques. Recent advances in large language models (LLMs) enable nuanced analysis of social media data, capturing linguistic patterns, contextual sentiment, and evolving themes \cite{10435239, tornberg2024large}. Integrating LLM-based approaches with statistical methods enhances both the precision and depth of our analysis. Guided by these methodological tools, our study addresses the following research questions:

\begin{itemize}
    \item[\textbf{RQ1:}] Descriptively, how does the public’s sentiment toward Donald Trump change over time and across different regions?
    \item[\textbf{RQ2:}] Does the assassination attempt exert a significant effect on public attitudes toward Donald Trump, after accounting for existing political divisions?
    \item[\textbf{RQ3:}] What are the major discussion topics before and after the assassination attempt, and how do these themes' sentiments evolve in response to the crisis? 
\end{itemize}


To answer these questions, we conduct a study of public discourse on X (formerly Twitter) before and after the July 2024 attempted assassination of Donald Trump. We first construct a prompt for aspect-based sentiment classification towards Trump, evaluate it on several contemporary models, and use the best-performing model to annotate all tweets in our dataset. We then employ a Difference-in-Differences (Did) approach to compare sentiment changes across different sets of states (e.g., Red vs. Blue, Swing vs. non-Swing) to assess the impact of the event on public sentiment. Subsequently, we apply LLM-based topic modeling to explore fine-grained changes in public discourse. Our findings reveal a general increase in positive sentiment toward Trump that is not significantly influenced by state-level partisanship. Moreover, we observe notable changes in the topics discussed, with a particularly sharp decline in tweets addressing controversies surrounding Trump after the event. These results align with a public response oriented towards sympathy, suggesting a broad improvement in sentiment toward Trump without evidence of heightened social tensions.

\section{Related Work}
The existing research on how political shocks, such as assassination attempts on public figures, influence public opinion can generally be divided into two main perspectives: sympathy and polarization.

The sympathy hypothesis indicates that significant public sympathy following a leader’s personal ordeal, rather than evaluations of their policy performance or partisan alignment, can result in a rapid and widespread increase in approval \cite{ostrom1985promise}. This theory is grounded in the broader framework of the ``Rally 'Round the Flag Effect'' \cite{brody1991assessing}, which identifies sharp spikes in presidential approval ratings during national crises, such as international conflicts or terrorist attacks \cite{mueller1970presidential, chanley2002trust, hetherington2003anatomy}. Over time, the ``Rally 'Round the Flag Effect'' has been extended beyond wartime scenarios to encompass various emergencies, including assassination attempts and even health crises involving presidents \cite{ostrom1985promise, brody1991assessing}. A compelling example is President Ronald Reagan, who saw a significant rise in approval ratings following his survival of an assassination attempt in 1981, a surge largely attributed to widespread public sympathy instead of his governance performance or political stance \cite{ostrom1985promise, brody1991assessing, gilbert2013attempted}. Based on the sympathy hypothesis, the assassination attempt on Donald Trump could trigger a rapid, broad-based increase in public sentiment, transcending existing political divisions.

The polarization hypothesis suggests that major social events, whether crises or shocks, can exacerbate political and ideological divisions as individuals react with divergent opinions and priorities \cite{putnam2007e}. This concept draws from Social Identity Theory, which asserts that group competition or crises intensify in-group loyalty and positive self-evaluations while fostering negative perceptions of out-groups \cite{tajfel2004social, huddy2003group}. For instance, \citet{sokol2018heavens} explores the assassination of Martin Luther King Jr., showing how the event heightened racial tensions and radicalized both supporters and opponents of the civil rights movement. If polarization is the predominant response, the assassination attempt on Donald Trump could intensify partisan divides, for example by exhibiting differential responses in ``red" versus ``blue" regions of the country. 



Previous scholarship has underscored the value of social media data in revealing how public sentiment unfolds in times of crisis, such as a national security threat or a health emergency \cite{bollen2010twitter, conover2011political, johansson2021rally}. For example, \citet{han2019using} employed a mixed-methods strategy combining qualitative content analysis with quantitative metrics to map social media interactions across platforms in a crisis scenario. Their study demonstrates how public sentiment can be more accurately captured by identifying specific discussion themes, analyzing user roles within the network, and integrating location-based data. 

Recent advancements in NLP, particularly LLMs, have improved the ability to extract insights from social media content \cite{dash2023sustainable, mahowald2024dissociating}. These models have become increasingly capable of identifying sarcasm and irony by leveraging contextual cues and learned patterns \cite{boutsikaris2024comparative}. They also excel in recognizing metaphorical expressions and implicit sentiment, in some cases approaching human-level performance in detecting subtle emotional undertones \cite{ziems2024can}. In addition, topic modeling offers a systematic approach to identifying emergent and dominant conversation themes, shedding light on how online users discuss impactful events \cite{blei2012probabilistic}. For instance, a recently developed tool called BERTopic, which uses BERT-based embeddings, has been applied to various computational social science tasks, including classifying tweets from disaster-affected regions into relevant topics for emergency management \cite{rachel2024topic} and identifying citizen concerns on urban governance through Reddit comments \cite{10591750}.


Moreover, integrating LLM-based techniques with causal inference (econometrics) tools offers a critical opportunity to deepen and refine analytical insights. Causal models such as DiD have been widely used to assess the causal impact of ``interventions'' (e.g. policies, or events) by comparing outcomes between treatment and control groups \cite{basu2017evaluating, wing2018designing, hsu2022increased}. For example, \citet{ferraz2008exposing} demonstrated how corruption disclosures influence electoral accountability using DiD. Similarly, \citet{holliday2024trump} employed the related method of Interrupted Time Series Analysis to examine how assassination attempts influenced voter attitudes toward both their own and opposing parties. 


These prior studies provide a robust methodological foundation, particularly the integration of LLM-based techniques with causal inference models, for investigating the impact of the political shock—the July 2024 Trump assassination attempt. In this case, this research aims to determine whether the event generates a general increase in support, as proposed by the sympathy hypothesis, or exacerbates social tensions, as suggested by the polarization hypothesis, through a large-scale analysis of posts on X.


\section{Data and Methods}
\subsection{Data collection}



We collected data from X using Brandwatch, a social media analytics tool offering historical data retrieval. The keyword ``Trump'' was selected collaboratively by domain and technical experts to encompass all relevant posts. Data was gathered from July 7 to July 20, 2024, covering one week before and after the assassination attempt on July 13, 2024. This two-week timeframe was chosen to capture the short-term impact of the event while minimizing confounding factors from unrelated developments.

The dataset was restricted to posts geotagged within the United States and written in English to focus on U.S. audiences. Posts lacking state-level location metadata were excluded, resulting in a final dataset of 122,526 relevant posts. This approach ensured a focused and accurate analysis of the public discourse surrounding the event.

\subsection{Stance detection and validation}


We used LLMs to perform aspect-based sentiment analysis (ABSA) on X posts, focusing on sentiments toward Donald Trump. Each post was classified into one of three categories---positive, neutral, or negative---based on the sentiment expressed toward him. 


To validate the performance of different LLMs, we manually annotated 300 randomly selected posts. Three annotators independently labeled each post, and then we applied a majority voting strategy to determine the final sentiment label of a post. This high-quality, human-annotated dataset provided a benchmark for evaluating the accuracy and reliability of the LLMs.

\begin{figure*}[t]
    \centering
    \includegraphics[width=1\textwidth]{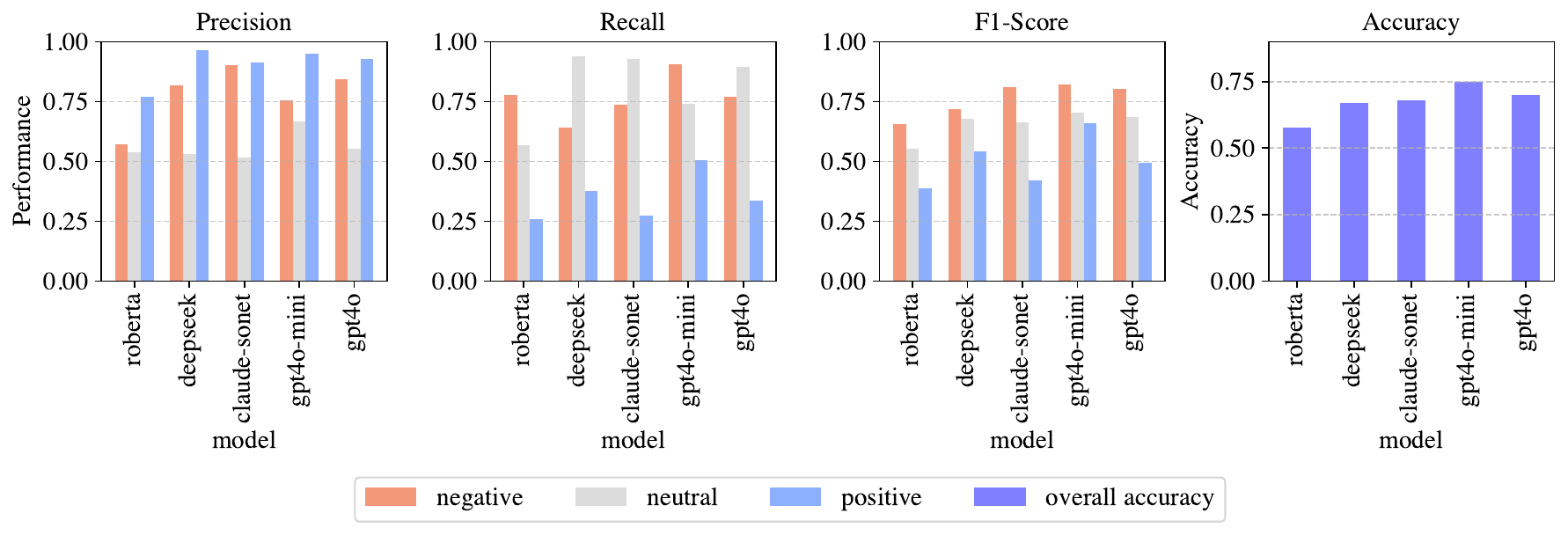}
    \caption{Performance evaluation of stance detection across different models.}
\label{fig:performance}
\end{figure*}

For the evaluation, we compared the performance of four LLMs, including DeepseekV2.5 \cite{deepseekai2024deepseekv2strongeconomicalefficient}, Claude-3.5-Sonnet-20240620 \cite{anthropic_claude_2024}, ChatGPT-4o-Latest \cite{gpt4o}, and GPT-4o-Mini \cite{gpt4o}, plus a conventional sentiment model called RoBERTa sentiment \cite{barbieri2020tweeteval}. A task-specific prompt was designed to optimize the models' performance, incorporating detailed definitions of the sentiment categories and explicit instructions for conducting aspect-based analysis. The full prompt is provided in Appendix. Each model was evaluated across four key performance metrics---precision, recall, F1-score, and overall accuracy---for each sentiment category (positive, neutral, and negative). The results, summarized in Figure~\ref{fig:performance}, illustrate the relative strengths and weaknesses of each model.

As shown in Figure~\ref{fig:performance}, GPT-4o-Mini consistently outperformed the other models in terms of accuracy and F1-scores across all sentiment categories. Its ability to capture nuanced expressions of sentiment, combined with its computational efficiency, made it the most suitable choice for scaling up the analysis to the full dataset. This result also highlights the significance of leveraging LLMs with prompt design over traditional sentiment analysis tools for detecting aspect-based sentiment from social media posts.


\subsection{Difference-in-differences (DiD) modeling}

We then employed the DiD approach---a robust quasi-experimental econometric method \cite{card2000minimum}---to estimate the impact of the assassination attempt on public sentiments. This study is particularly well-suited for the DiD approach as it examines the causal effect of a clearly defined intervention—the assassination attempt—on a measurable outcome, public sentiment. Also, the natural separation of states into treatment and control groups based on partisan alignment provides the necessary conditions to apply the DiD framework effectively. 

Implementing the DiD models typically involves three steps \cite{besley2002unnatural, ladd2009exploiting, angrist2009mostly}: (1) verifying that the treatment and control groups meet the parallel trends assumption in the absence of the intervention, (2) constructing regression models to estimate the treatment effect, and (3) conducting robustness checks, such as placebo tests, to validate the reliability of the results.

\textbf{\textit{Intervention point specification.}} In this study, the intervention refers to the assassination attempt on Donald Trump, which occurred at 6:11 PM Eastern Daylight Time (EDT) on July 13, 2024 \cite{odonoghue2024secret}. To ensure consistency across time zones and accurately assess the impact of this event on public attitudes nationwide, we converted all timestamps to Coordinated Universal Time (UTC), which is EDT+4. The data was then divided into two periods based on UTC: before the intervention (July 7–July 13, up to 10:11 PM UTC) and after the intervention (July 13, starting at 10:11 PM UTC–July 20). The public sentiment score toward Donald Trump of each state and date was calculated using annotations generated by the GPT-4o-mini model. To quantify public attitudes, the data was aggregated by state and date, with a weighted average sentiment score for X posts posted in a given state on a specific day, $S_{s,d}$:

\begin{equation}
S_{s,d} = \frac{\sum_{i=1}^{N_{s,d}} S_i}{N_{s,d}}
\end{equation}

where: $N_{s,d}$ is the total number of tweets posted in state $s$ on date $d$; $S_i$ is the sentiment score for the $i$-th tweet in state $s$ on date $d$.

\textbf{\textit{Treatment-Control group division.}} As mentioned above, the key distinction between the sympathy hypothesis and the polarization hypothesis lies in whether the assassination attempt leads to significantly different trends or magnitudes in public attitudes across social groups. Given the deep political polarization in the United States \cite{bennett2021mapping, rueda2023disaggregated}, we established three treatment-control group divisions based on state-level political preferences: (1) Republican-leaning (``red'') states as the treatment group, with Democrat-leaning (``blue'') states as the control group; (2) Swing states as the treatment group, with blue states as the control group; and (3) Swing states as the treatment group, with red states as the control group. This approach is grounded in the rationale that red states are likely to exhibit the strongest net increase in sentiment compared to non-red states. The classification of red, blue, and swing states is presented in the Table~\ref{tab:political_demarcation}, based on public polling data provided by Politico \cite{politico_swing_states_2024}. 


\begin{table*}[htbp] 
\centering
\setlength{\tabcolsep}{4pt} 
\small
\caption{Political Demarcation based on the Politico} 
\begin{tabular}{p{0.16\textwidth}p{0.8\textwidth}}
\toprule
\textbf{Category} & \textbf{States} \\
\midrule
Swing States & Arizona, Georgia, Michigan, Nevada, North Carolina, Pennsylvania, Wisconsin \\
\midrule
Blue States & California, Colorado, Connecticut, Delaware, Hawaii, Illinois, Maine, Maryland, Massachusetts, Minnesota, New Jersey, New Mexico, New York, Oregon, Rhode Island, Vermont, Washington, Virginia \\
\midrule
Red States & Alabama, Alaska, Arkansas, Idaho, Indiana, Iowa, Kansas, Kentucky, Louisiana, Mississippi, Missouri, Montana, Nebraska, North Dakota, Oklahoma, South Carolina, South Dakota, Tennessee, Texas, Utah, West Virginia, Wyoming, Florida, Ohio, New Hampshire \\
\bottomrule
\end{tabular}
\label{tab:political_demarcation}
\normalsize
\end{table*}

\textbf{\textit{Parallel trend verification.}} This assumption ensures that, in the absence of the intervention, the treatment and control groups would have followed similar trends over time. To test this, we conducted a linear regression analysis for each state individually using the following equation.


\begin{equation}
y_i = \beta_0 + \beta_1 x_i + \epsilon_i
\end{equation}

where: \( y_i \) is the public sentiment score for a given day; \( x_i \) is the number of days since the earliest observation in that state; \( \beta_1 \) is the slope, representing the rate of change in public sentiments over time.

To quantify the difference between the mean slopes of the two groups relative to the variability within each group, we performed an independent samples t-test analysis:

\begin{equation}
t = \frac{\bar{\beta}_T - \bar{\beta}_C}{\sqrt{\frac{s_T^2}{n_T} + \frac{s_C^2}{n_C}}}
\end{equation}

where: \( \bar{\beta}_T \) and \( \bar{\beta}_C \) are mean slopes for the treatment group and the control group, individually; \( s_T^2 \) and \( s_C^2 \) are the variance of slopes in the treatment group and the control group, individually; \( n_T \) and \( n_C \) are the number of states in the treatment group and the control group, individually.

The resulting p-values were used to evaluate the statistical significance of the t-statistic. As shown in Table~\ref{tab:parallel_trends}, the pre-intervention trends in public sentiments for all three sets of treatment-control groups are statistically similar (all p-values $>$ 0.10). This confirms that the parallel trends assumption holds, allowing us to proceed with constructing the DiD models for further analysis.

\begin{table}[htbp]
\centering
\caption{T-Test for Parallel Trends Assumption Verification}
\small
\begin{tabular}{p{0.32\textwidth}c}
\toprule
\textbf{Treatment v.s. Control Groups} & \textbf{T-Statistic} \\ \midrule
Red v.s. Blue States & 0.357 \\
Swing v.s. Red States & 0.075 \\
Swing v.s. Blue States & 0.703 \\ \bottomrule
\multicolumn{2}{l}{\footnotesize Significance codes: * p $<$ 0.05, ** p $<$ 0.01, *** p $<$ 0.001} \\
\end{tabular}
\label{tab:parallel_trends}
\end{table}

\textbf{\textit{Estimation framework.}} To estimate the causal effect of the intervention on public sentiments toward Donald Trump, we specified an Ordinary Least Squares (OLS) regression model incorporating key covariates and interaction terms:
\begin{equation}
\begin{aligned}
 Sentiment\ Score = & \ \beta_0 + \beta_1(Group) + \beta_2(Event) \\
& + \beta_3(Group \times Event) \\
& + Controls + \epsilon
\end{aligned}
\end{equation}

where:
\begin{itemize}
    \item \( Sentiment\ Score \): The dependent variable representing the weighted average public attitude score toward Donald Trump.
    \item \( Group \): A binary variable indicating the treatment group.
    \item \( Event \): A binary variable indicating whether the observation occurs before or after the intervention.
    \item \( Group \times Event \): An interaction term capturing the differential effect of the intervention on the treatment group.
    \item \( Controls \): A set of standardized control variables, expressed as, including median income, and the percentage of individuals with a bachelor’s degree in certain states, and a standardized temporal variable to account for time trends.
    \item \( \epsilon \): The error term.
\end{itemize}

This framework enables a rigorous examination of the intervention's impact by isolating the treatment effect (\( \beta_3 \)) while controlling for regional socioeconomic characteristics and temporal dynamics.

\textbf{\textit{Robustness check.}} To validate the robustness of the causal inference framework, we conducted a placebo test by setting the intervention break-point to July 11, 2024, around two days prior to the actual assassination attempt. An OLS regression model was also specified, incorporating key covariates and interaction terms:
\begin{equation}
\begin{aligned}
 Sentiment\ Score = & \ \beta_0 + \beta_1(Group) + \beta_2(Placebo) \\
& + \beta_3(Group \times Placebo) \\
& + Controls + \epsilon
\end{aligned}
\end{equation}

where:
\begin{itemize}
    \item \( Placebo \): A binary variable indicating whether the observation occurred before or after the placebo intervention on July 11, 2024.
    \item \( Group \times Placebo \): An interaction term capturing the differential effect of the placebo intervention on the treatment group.
\end{itemize}

This placebo test can provide a robustness check to ensure that spurious relationships or unaccounted temporal trends do not drive any observed effects in the primary analysis. 

\subsection{Topic modeling}
Topic modeling is a widely used method for uncovering semantic themes in large text corpora, including social media posts \cite{li2024chatgpt} and news articles \cite{xian2024landscape}. For our analysis of the X dataset, we employed the BERTopic framework \cite{grootendorst2022bertopic}, which leverages text embeddings to capture nuanced, context-aware representations. Unlike traditional approaches such as Latent Dirichlet Allocation (LDA) \cite{Jelodar2019}, BERTopic integrates contextual word meanings, offering greater accuracy for complex topics like public discourse on the Trump assassination attempt. We used the \texttt{all-MiniLM-L6-v2} embedding model from Sentence Transformers \cite{reimers-2019-sentence-bert} for its balance between efficiency and performance.

To refine the high-dimensional BERT embeddings, we employed Uniform Manifold Approximation and Projection (UMAP) for dimensionality reduction, enhancing clustering and interpretability \cite{mcinnes2018umap}. Initial modeling identified over 200 topics, which were reduced to 50 clusters using the elbow method \cite{marutho2018determination}, as illustrated in Figure~\ref{fig:elbow} in the Appendix.

To analyze temporal nuances, we fit separate topic models for posts before and after the assassination attempt. Clusters were represented using class-based Term Frequency–Inverse Document Frequency (c-TF-IDF) \cite{grootendorst2022bertopic}, which identified key terms and documents. For further abstraction, GPT-4o \cite{gpt4o} consolidated the 100 resulting topics into broader umbrella topics based on representative words. To ensure accuracy, three annotators reviewed and validated the categorization using a majority voting strategy. This multi-step process distilled complex discussions into structured, high-level themes while preserving contextual depth.

\section{Results}
\subsection{RQ1: Descriptively, how does the public’s sentiment toward Donald Trump fluctuate over time and across different regions?}
\begin{figure}[htbp]
    \centering
    \includegraphics[width=0.45\textwidth]{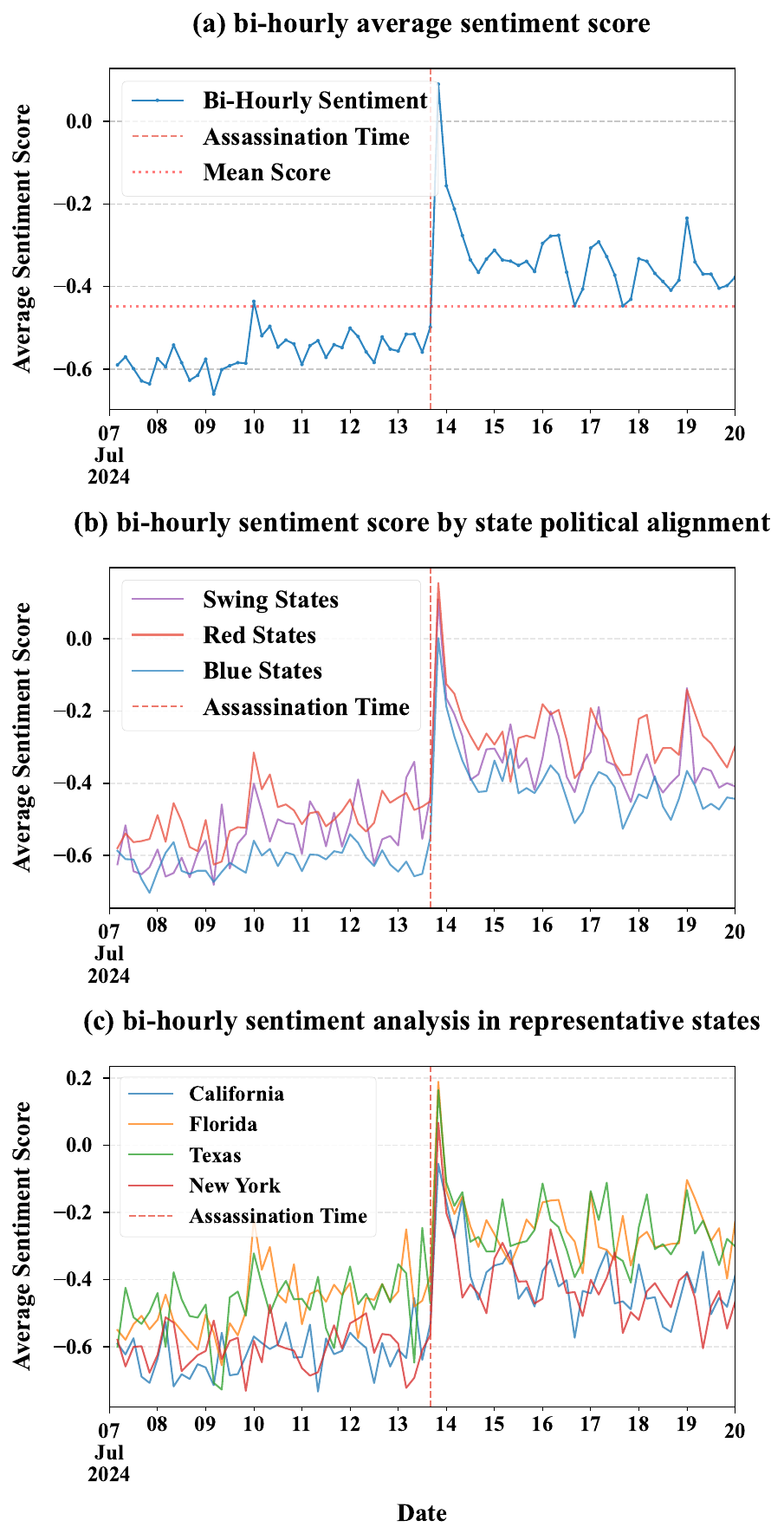}
    \caption{Temporal evolution of public sentiment following the attempted assassination.}
\label{fig:trend}
\end{figure}

Analysis of X's sentiment toward Donald Trump reveals a distinct temporal shift surrounding the July 13, 2024, assassination attempt (Figure~\ref{fig:trend}a). Before July 13, sentiment scores average around $-0.55$. On the day of the event, these scores rise sharply to approximately $-0.20$, indicating a significant but temporary reduction in negativity. In the days following, the sentiment scores declines but stabilizes at about $-0.35$, suggesting a modest and persistent improvement compared to pre-event levels.

Disaggregating sentiment by political alignment reveals consistent trends across all groups, with no pronounced regional outliers (Figure~\ref{fig:trend}b). Red States exhibited slightly higher sentiment (around $-0.45$) compared to Blue States (approximately $-0.60$) before the event. Following the incident, both groups experienced noticeable positive shifts, with Red States peaking at $-0.10$ and Blue States reaching $-0.32$. Swing States followed a similar trajectory, reflecting a broadly consistent response across regions, regardless of ideological leanings. State-level sentiment trends (Figure~\ref{fig:trend}c) for key states like Texas, Florida, California, and New York further confirm this pattern, with all four experiencing comparable shifts, albeit with minor variations in magnitude.

\begin{figure}[htbp]
    \centering
    \includegraphics[width=0.45\textwidth]{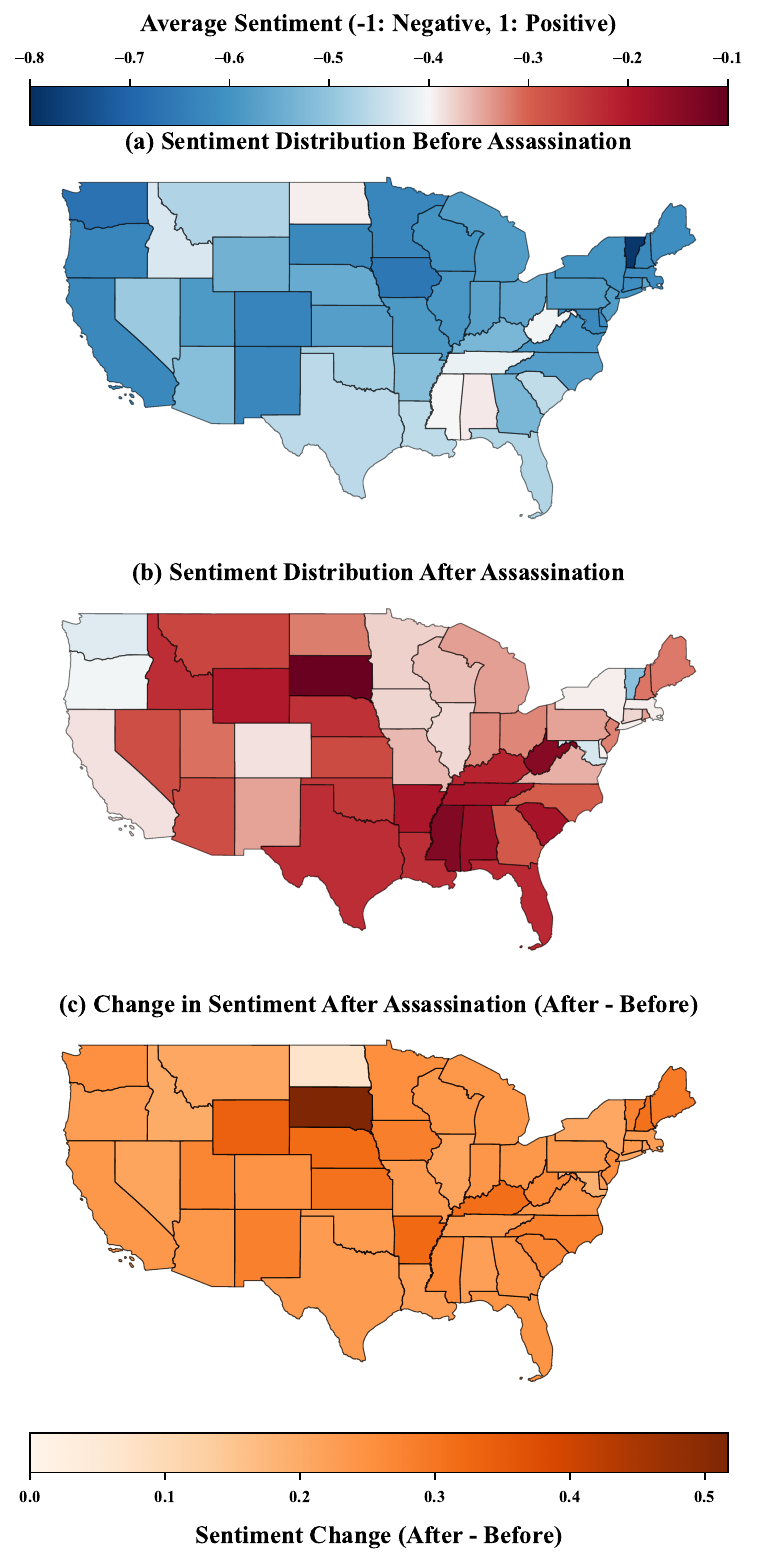}
    \caption{Geographic distribution of public sentiment before and after the assassination.}
\label{fig:gis}
\end{figure}

To further explore these patterns, we examined the geographic distribution of sentiment changes across all states, as illustrated in Figures~\ref{fig:gis}a,\ref{fig:gis}b, and \ref{fig:gis}c. This analysis revealed uniform sentiment improvements nationwide, with states that were previously more negative, such as North Dakota and Nebraska, showing shifts similar to those in states like California and New York. Despite minor variations in magnitude, the absence of significant regional outliers suggests that the assassination attempt elicited a broadly consistent response across the United States (as shown in the Figure~\ref{fig:gis} in Appendix). These geographical findings reinforce the state-level sentiment trends and provide a cohesive picture of how public attitudes shifted uniformly, transcending partisan and regional boundaries.


In addition to the sentiment trend analysis, we conducted correlation analyses to explore the relationships between socioeconomic and demographic factors and public sentiment before and after the assassination attempt. The results of this analysis, presented in Appendix, highlight patterns such as higher positive sentiment in Republican-leaning states and more negative sentiment in areas with higher educational attainment or income levels. While these findings offer valuable context, their interpretation is limited by the small sample size of 50 states and thus should be considered exploratory. For a detailed discussion and table of results, please refer to Appendix.


\subsection{RQ2: Does the assassination attempt exert a significant effect on public sentiments toward Donald Trump, after accounting for existing political divisions?}
In this section, we continue to investigate the impact of the event on public attitudes by building a causal inference model. While RQ1 provides a descriptive overview of sentiment trends and their variation, it does not isolate the specific effect of the assassination attempt from broader temporal fluctuations or other concurrent factors. To address this, we employ the DiD modeling to estimate the causal effect of the event by comparing changes in sentiment across treated and control groups before and after the event.

\textbf{\textit{The major estimation framework.}} The major DiD estimation results are presented in the Table~\ref{tab:major_estimation}, providing insights into the impact of the assassination attempt on public attitudes across different treatment and control group comparisons. In particular, Model (1) defines red states as the treatment group and blue states as the control group. Model (2) compares swing states as the treatment group against red states as the control group, while Model (3) compares swing states as the treatment group against blue states as the control group. Across all three models, the dependent variable is the weighted public sentiment score toward Donald Trump. 

Across all three models, the \( Event \) coefficients are positive and statistically significant, indicating a general increase in public sentiment toward Trump after the assassination attempt. Nevertheless, the \( Group\_Event \) interaction term is not statistically significant in any of the models. This suggests that the event did not lead to a disproportionately greater increase in public support for Trump in red states compared to blue states, nor in swing states compared to either red or blue states.

These findings indicate that the assassination attempt did not trigger party affective polarization, as no group exhibited a significantly stronger shift in public attitudes. Instead, the observed rise in support for Trump across all groups aligns more closely with the sympathy hypothesis, where public sentiment improves broadly in response to a leader's personal ordeal.

\begin{table*}[htbp] 
\centering
\caption{DiD Major Estimation Results: Treatment vs. Control Groups}
\label{tab:major_estimation}
\small 
\renewcommand{\arraystretch}{1.2} 
\setlength{\tabcolsep}{1pt} 
\begin{tabular}{p{0.2\linewidth} >{\centering\arraybackslash}p{0.23\linewidth} >{\centering\arraybackslash}p{0.23\linewidth} >{\centering\arraybackslash}p{0.23\linewidth}} 
\hline
\multicolumn{4}{p{0.8\linewidth}}{\textbf{Dependent Variable:} Averaged Public Attitude Scores Toward Donald Trump} \\ 
 & \textbf{(1) Red vs. Blue States} & \textbf{(2) Swing vs. Red States} & \textbf{(3) Swing vs. Blue States} \\
\hline
Intercept & -0.695 (0.024)** & -0.577 (0.022)** & -0.730 (0.021)** \\
Group & 0.090 (0.027)** & 0.045 (0.034) & 0.063 (0.031)* \\
Event & 0.418 (0.036)** & 0.414 (0.036)** & 0.456 (0.033)** \\
Group\_Event & 0.026 (0.034) & -0.008 (0.045) & 0.018 (0.038) \\
Controls & Yes & Yes & Yes \\
No. Observations & 613 & 463 & 360 \\
Adj. R-Squared & 0.380 & 0.355 & 0.451 \\
\hline
\end{tabular}
\caption*{Significance codes: * p $<$ 0.05, ** p $<$ 0.01, *** p $<$ 0.001}
\end{table*}

\textbf{\textit{Robustness check.}} For the robustness check, a placebo test was conducted by shifting the intervention date to July 11, 2024, two days prior to the actual assassination attempt. This analysis evaluates whether the observed effects in the main estimation framework were due to random noise or pre-existing trends rather than the event itself. The results of the placebo test are presented in Table~\ref{tab:placebo_test}. 


Across all three placebo models, the \( Group\_Placebo \) interaction term, which measures the differential impact of the placebo intervention on the treatment groups compared to the control groups, is not statistically significant in any model. In addition, the \( Placebo \) coefficients, capturing the overall change in public attitudes after the placebo date, are positive and statistically significant, indicating a general upward trend in attitudes toward Trump even before the actual assassination attempt. While all the \( Placebo \) coefficients are significant, their magnitudes are notably smaller than the \( Event \) coefficients in the main models.



By integrating the findings from the main estimation framework and the placebo test, it is fairly to conclude that the assassination attempt is the primary catalyst for the observed sympathy hypothesis. Moreover, the absence of statistically significant interaction effects in both the main and placebo models implies that the event does not lead to social polarization since no group exhibits a disproportionate change in public sentiments compared to others. Instead, the general increase in public sentiment aligns with the sympathy hypothesis, where public attitudes improve broadly in response to a critical event, rather than being driven by existing partisan-specific dynamics.

\begin{table*}[htbp] 
\centering
\caption{DiD Placebo Test: Treatment vs. Control Groups}
\label{tab:placebo_test}
\small 
\renewcommand{\arraystretch}{1.2} 
\setlength{\tabcolsep}{1pt}
\begin{tabular}{p{0.2\linewidth} >{\centering\arraybackslash}p{0.23\linewidth} >{\centering\arraybackslash}p{0.23\linewidth} >{\centering\arraybackslash}p{0.23\linewidth}} 
\hline
\multicolumn{4}{p{0.8\linewidth}}{\textbf{Dependent Variable:} Averaged Public Attitude Scores Toward Donald Trump} \\ 
 & \textbf{(1) Red vs. Blue States} & \textbf{(2) Swing vs. Red States} & \textbf{(3) Swing vs. Blue States} \\
\hline
Intercept & -0.660 (0.031)** & -0.550 (0.028)** & -0.690 (0.029)** \\
Group & 0.081 (0.034)* & -0.032 (0.043) & 0.066 (0.042) \\
Placebo & 0.283 (0.039)** & 0.291 (0.039)** & 0.304 (0.039)** \\
Group\_Placebo & 0.036 (0.038) & -0.025 (0.052) & 0.011 (0.047) \\
Controls & Yes & Yes & Yes \\
No. Observations & 613 & 463 & 360 \\
Adj. R-Squared & 0.272 & 0.251 & 0.265 \\
\hline
\end{tabular}
\end{table*}

\subsection{RQ3: What are the major discussion topics before and after the
assassination attempt, and how do these themes’ sentiments evolve in response to the crisis?}
\textbf{What are people talking on social media?} In this subsection, we described our general findings from topic modeling. In Table~\ref{tab:topicmodeling}, we demonstrate our 10 umbrella topics, two of their sub-topics, and representative documents.

\begin{table*}[htbp]
\small
\centering
\caption{Topic Modeling Results and Their Representative Documents}
\begin{tabular}{p{2.5cm}p{1.5cm}p{4.5cm}p{8cm}}
\toprule
\textbf{Umbrella Topics} & \textbf{Posts} & \textbf{Representative Subtopics} & \textbf{Representative Documents} \\
\midrule
MAGA & 20,377 & maga, medicare, repeal,... \newline maga, donaldtrump, patriot... & ``if social security be important to you vote for trump voting for project'' 2025... \newline ``maga trump'' \\
\addlinespace[0.25em]
Assassinations & 23,330 & shoot, shooting, shot,... \newline assassinat, conspiracy, assassinat... & ``this be how i find out donald trump get shoot'' \newline ``this be why trump need to go after the deep state up on day'' \\
\addlinespace[0.25em]
General Voting and Elections & 16,458 & vote, voting, poll,.. \newline president, presidenti, presidency, ... & ``a vote for joe biden be a vote for war a vote for president trump be a vote for peace'' \newline ``if trump become president again in it will be the final death blow to the global order'' \\
\addlinespace[0.25em]
Media & 4,900 & truth, post, tweet,... \newline msnbc, media, msm,... & ``trump just post this on truth'' \newline ``a side of president trump that the mainstream medium do want you to see'' \\
\addlinespace[0.25em]
Democrats in Elections & 11,954 & dems, democrats, democrat,... \newline pelosi, assassinat, psaki,... & ``democrats and joe be try can only do one thing in their last attempt to rid trump...'' \newline ``paul pelosi be almost assassinate and trump mock...'' \\
\addlinespace[0.25em]
Republicans in Elections & 6,265 & pence, trumper, vice,... \newline haley, rnc, nikki,... & ``trump pick my senator jd vance of ohio for his vp pick...'' \newline nikki haley voter if trump be the republican nominee... \\
\addlinespace[0.25em]
Trump Controversies & 21,723 & hitler, adolf, goebbels,... \newline threaten, threat, dems,... & ``donald trump be not hitler'' \newline ``his be just one example biden say donald trump be a genuine threat to this nation'' \\
\addlinespace[0.25em]
Race and Social Issues & 3,228 & blacks, black, n***a,... \newline deport, deportatio, immigrants,... & ``if you be black and vote for trump then you not a real black'' \newline ``donald trump on illegal immigrant this be poison our country'' \\
\addlinespace[0.25em]
International Politics & 2,777 & putin, russia, russias,... \newline israel, isreal, netanyahu,... & ``you back trump you back russia'' \newline ``trump will give israel everything they ask for and more'' \\
\addlinespace[0.25em]
Pray for Trump & 2,924 & pray, prayers, prayer,... \newline bless, blessed, god & ``pray for trump'' \newline ``god bless our president trump'' \\
\addlinespace[0.25em]
Unrelated & 8,590 & bitch, shit, shill... & ``the woman say herself trump do call'' \\
\bottomrule
\end{tabular}
\label{tab:topicmodeling}
\end{table*}

In our topic modeling results, key themes identified include a notable emphasis on conspiracies and concerns about potential violence, captured under the ``Assassinations'' umbrella topic. This theme encompasses discourse around threats, shootings, and related incidents, reflecting heightened public attention to this event. Other related topics highlight polarized views, with some discussing broader implications for the electoral process, media narratives, and international politics about Trump. Additionally, conversations under ``Trump Controversies'' and ``Media'' indicate significant public engagement, often critiquing political rivals, media framing, or broader societal impacts. These results provide insights into the complex and emotionally charged nature of public discourse during this period of heightened concern.

\textbf{How do topics of interest change before and after the assassination attempt?} 
\begin{figure}[htbp]
    \centering
    \includegraphics[width=\columnwidth]{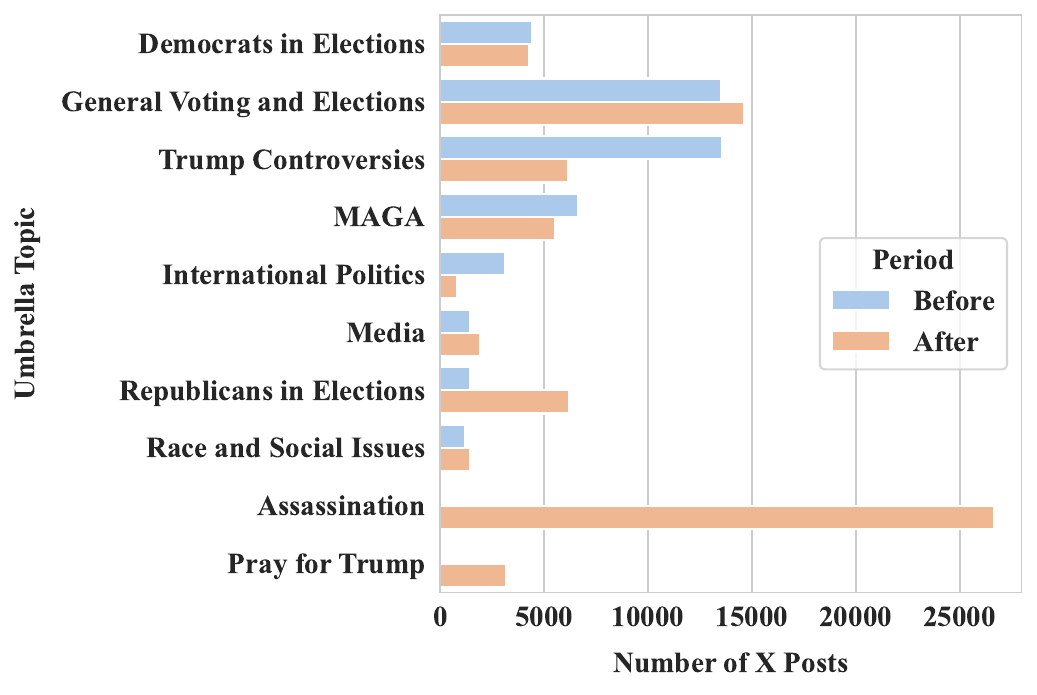}
    \vspace{-5pt}
    \caption{Changes in the frequency of broad topic categories before vs. after the assassination attempt}
\label{fig:topic_distribution}
\end{figure}

Figure~\ref{fig:topic_distribution} displays the number of topic changes after the assassination attempt. This result implies significant shifts in the thematic focus of posts before and after the assassination event. Before the assassination attempt, the primary discussion areas were `Trump Controversies' and `General Voting and Elections,' both of which show comparable levels of engagement, indicating the public's active involvement in and attention to broader electoral processes and controversies surrounding Trump. Topics such as `MAGA,' `International Politics,' and `Race and Social Issues' also attracted notable, though less substantial, attention, reflecting diverse yet secondary concerns among users at the time. After the assassination attempt, we observe a dramatic surge in posts categorized under `Assassination' and `Pray for Trump,' with `Assassination' becoming the overwhelmingly dominant topic. These two categories vastly outpace others in terms of post volume, indicating a significant redirection of public discourse and a collective emotional and informational response to the event.

Interestingly, while most other themes, such as `Democrats in Elections' and `MAGA,' remain relatively stable, other umbrella topics exhibit notable changes. For instance, `Trump Controversies' displays a marked decline, suggesting that the assassination attempt may have temporarily muted critical discourse and shifted the tone of conversations away from negativity. This decline could reflect a broader social tendency to show solidarity or refrain from criticism during moments of national or personal crisis. Conversely, we also observe a noticeable rise in posts categorized under `Republicans in Elections,' which coincides with discussions about Trump’s announcement of J.D. Vance as his vice-presidential nominee on July 15, 2024. This surge highlights the potential interplay between significant political developments and unexpected events, where the latter may amplify or recontextualize political discourse.

Overall, these results show how online discourse patterns shift following the assassination attempt.. The patterns indicate a clear change in topic focus, with increased attention to the assassination event and supportive messages, while previously dominant topics like `Controversies' receive less attention. This suggests that significant political shock events like assassination attempt can temporarily redirect public discussion priorities on social media platforms. Again, this observation aligns with the sympathy hypothesis with expressions of emotional support (e.g., `Pray for Trump') taking precedence in the online discourse.

\begin{figure*}[htbp]
    \centering
    \includegraphics[width=\textwidth]{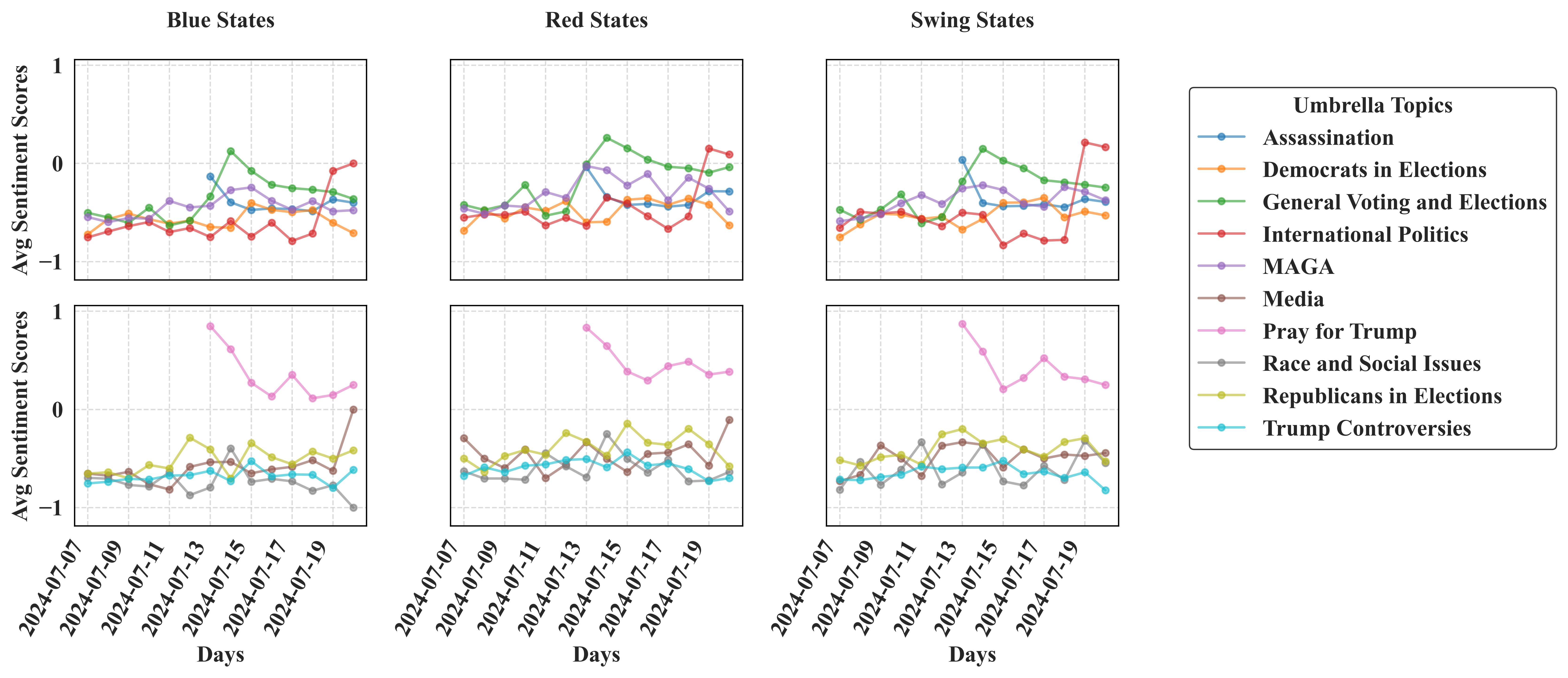}
    \caption{Sentiment scores by topics and types of states along time}
\label{fig:topic_sentiment}
\end{figure*}

\textbf{What are the sentiment scores towards different topics among blue/red/swing states?} We visualized the average sentiment scores along time in three types of states, as shown in Figure~\ref{fig:topic_sentiment}. 

A prominent trend is the sharp surge in positive sentiment for the ``Pray for Trump'' topic, peaking around July 15, particularly in red states. This increase reflects widespread emotional support and solidarity in the aftermath of the assassination, especially among Trump supporters, who may view the event as a personal or political attack on their leader. The spike also coincides with increased religiosity in online discourse, with ``prayer'' acting as a unifying theme across diverse political alignments, albeit to varying degrees.

Conversely, the sentiment for ``Assassination'' shows an initial peak but declines steadily over time. This pattern likely reflects an emotional progression from immediate shock and grief toward more critical or factual discussions as the public processes the event. In red states, the decline in ``Assassination'' sentiment is slower, possibly due to ongoing discussions about its implications for Trump and his political base. Meanwhile, topics such as ``General Voting and Elections'' and ``Republicans in Elections'' also experience notable fluctuations, with heightened positive sentiment around July 15 aligning with Trump's announcement of J.D. Vance as his vice-presidential nominee. This suggests that the political momentum generated by the nomination temporarily bolstered optimism among Republicans, especially in red states.

Interestingly, ``Trump Controversies'' remains consistently negative across all states, but its overall prominence decreases post-assassination, likely reflecting a shift in focus from criticism to solidarity during a moment of crisis. Red states exhibit higher positive sentiment in both ``Pray for Trump'' and ``Republicans in Elections,'' compared to blue and swing states, indicating a stronger alignment with Trump and a more emotionally supportive response. In contrast, blue states maintain more neutral or negative tones in these topics, reflecting partisan divides in the interpretation of and reaction to the assassination. These findings highlight how key events, such as an assassination, amplify emotional responses, shift the tone of discourse, and emphasize regional differences in political and social alignment.

\section{Discussion}





The findings of this study reveal the nuanced public response to the July 2024 assassination attempt on Donald Trump, with key insights emerging from sentiment analysis, causal modeling, and topic modeling. Our results indicate a significant yet uniform increase in public sentiment across states and political affiliations, suggesting the event elicited broad empathy and a reduction in negative sentiment. This aligns with the sympathy hypothesis, where acts of violence against prominent figures often generate public support and solidarity. Notably, DiD analysis confirmed that the incident had a unifying impact, as the absence of significant interaction effects demonstrates that it did not exacerbate existing ideological divides.

These findings resonate with \citet{holliday2024trump}, who similarly observed a reduction in support for partisan violence and an increase in in-group attachment among Republicans, particularly MAGA Republicans, following the assassination attempt. However, while Holliday et al. focus on decreased hostility and political violence, our study emphasizes the broader sentiment shifts across political and geographical lines, revealing a nationwide pattern of reduced negativity. Additionally, Holliday et al. noted the persistence of affective polarization, with no significant change in out-group hostility among Democrats or Republicans, a nuance consistent with our observation that sentiment improvements were largely non-polarizing.

Our topic modeling analysis complements these results by revealing significant shifts in public discourse. Before the assassination attempt, discussions were dominated by themes such as ``Trump Controversies" and ``General Voting and Election.'' In the aftermath, the discourse shifted toward less contentious topics, such as ``Prayers for Trump'' and expressions of concern for his safety in ``Assassination''. These changes underscore how high-profile political crises can temporarily redirect public focus from divisive narratives to themes reflecting personal or emotional engagement. Together, our findings and those of \citet{holliday2024trump} illustrate the complex interplay between sentiment, partisanship, and discourse during moments of political crisis, highlighting both the unifying and the nuanced emotional dynamics such events can evoke.

\subsection{Future Works}

This study presents several avenues for future research. The first direction is related to data resource and collection. While the Brandwatch API provides a sampling of data sufficient for our analysis, it may overlook certain information or insights from social media discourse. Additionally, X data might overrepresent certain user groups \cite{blank2017digital}, although political leanings may have shifted after Elon Musk's acquisition of the platform \cite{robertson2023here}. Future work could explore combining data from other social media platforms or news outlets to expand our analysis, aiming to generate a more representative view of public discourse.

Another area for future exploration involves extending the time frame of our analysis. We limit our study to a two-week period to minimize confounding factors, as public sentiment regarding Trump could be easily influenced by other events. While this short-term focus helps focus the effects of the assassination, an extended analysis could provide deeper insights into how social media sentiment evolves over time and how it aligns with real-world political events.

Last, future research could benefit from the application of more advanced NLP techniques. For example, while our aspect-based sentiment analysis involves designing specific prompts and testing various LLMs, there remains room for improvement. Future studies could employ advanced methods like Retrieval-Augmented Generation (RAG) \cite{gao2023retrieval} to refine sentiment classification. Additionally, for the BERTopic modeling, testing alternative embeddings, such as GPT embeddings, could probably enhance the coherence and interpretability of the topic clusters.

\section{Acknowledgment}
We would like to thank Libby Hemphill for her valuable feedback, which helps to improve this paper, and Lizhou Fan for assisting us with data annotation. We also want to extend our gratitude to the Linguistic Mechanisms Lab at Northwestern University for their insightful comments, which further enhance the quality of this work.

\bibliography{aaai22}

\newpage
\section*{Appendix}
\subsection{Prompt Template for Sentiment Analysis}
\label{appendix:prompt}
The following shows the prompt design for the sentiment classification given a X post. 
\begin{tcolorbox}[colback=gray!10!white, colframe=gray!50!gray, halign=left, top=0mm, bottom=0mm, left=1mm, right=1mm, boxrule=0.5pt]
\fontsize{9pt}{9pt}\selectfont 
\linespread{1}\selectfont
You are an expert in natural language processing and sentiment analysis. Your task is to analyze tweets about Donald Trump and identify the associated sentiment. \\
\vspace{10pt}
Background: On July 13, 2024, Donald Trump, a former president of the United States and then the presumptive nominee of the Republican Party in the 2024 presidential election, survived an assassination attempt while speaking at an open-air campaign rally near Butler, Pennsylvania. Trump was shot and wounded in his upper right ear by Thomas Matthew Crooks, a 20-year-old man from Bethel Park, Pennsylvania, who fired eight rounds from an AR-15--style rifle from the roof of a nearby building. \\
\vspace{10pt}
Task: Here is one tweet about Trump related to the event above. Please classify the sentiment of each tweet into one of the following categories: positive, negative, or neutral. \\
\vspace{10pt}
Instructions: \\
\begin{itemize}
    \item The definition of positive sentiments: Expresses approval, support, or praise for Trump, his actions, or his policies; Shows empathy toward Trump (e.g., well-wishing or prayers for recovery); Uses positive language, compliments, or expressions of agreement; Highlights perceived successes, achievements, or positive outcomes of his actions.
    \item The definition of negative sentiments: Expresses disapproval, criticism, or opposition to Trump, his actions, or policies. Uses negative language, insults, or expressions of disagreement. Highlights perceived failures, negative consequences, or harmful outcomes of his actions.
    \item The definition of neutral sentiments: Presents mixed or conflicting sentiments, balancing both positive and negative aspects without leaning toward either side. Discusses topics unrelated to Trump's persona, actions, or policies (e.g., comments on his supporters).
    \item You should only consider the person Donald Trump. In some tweets, there might be sentiments toward his relatives or his supporters. Please ignore those sentiments and treat them as neutral.
    \item Some tweets might include multiple sentiments. Please focus on the overall sentiment toward Trump and choose the most dominant sentiment.
\end{itemize}
\vspace{10pt}
Formatting: 
\begin{itemize}
    \item Please only provide a label for each tweet (positive, negative, or neutral).
    \item Do not include any additional information or context.
    \item You will be given a great penalty if you provide redundant information or context.
\end{itemize}
\vspace{10pt}
Demonstration:
If the tweet is: ``I love Trump! He is the best president ever!''
Your output should be: ``positive''. \\
\vspace{10pt}
Now please classify the sentiment of the following the tweet.
\end{tcolorbox}

\subsection{Topic Modeling Elbow Method Graph}
\begin{figure}[htbp]
    \centering
    \includegraphics[width=\columnwidth]{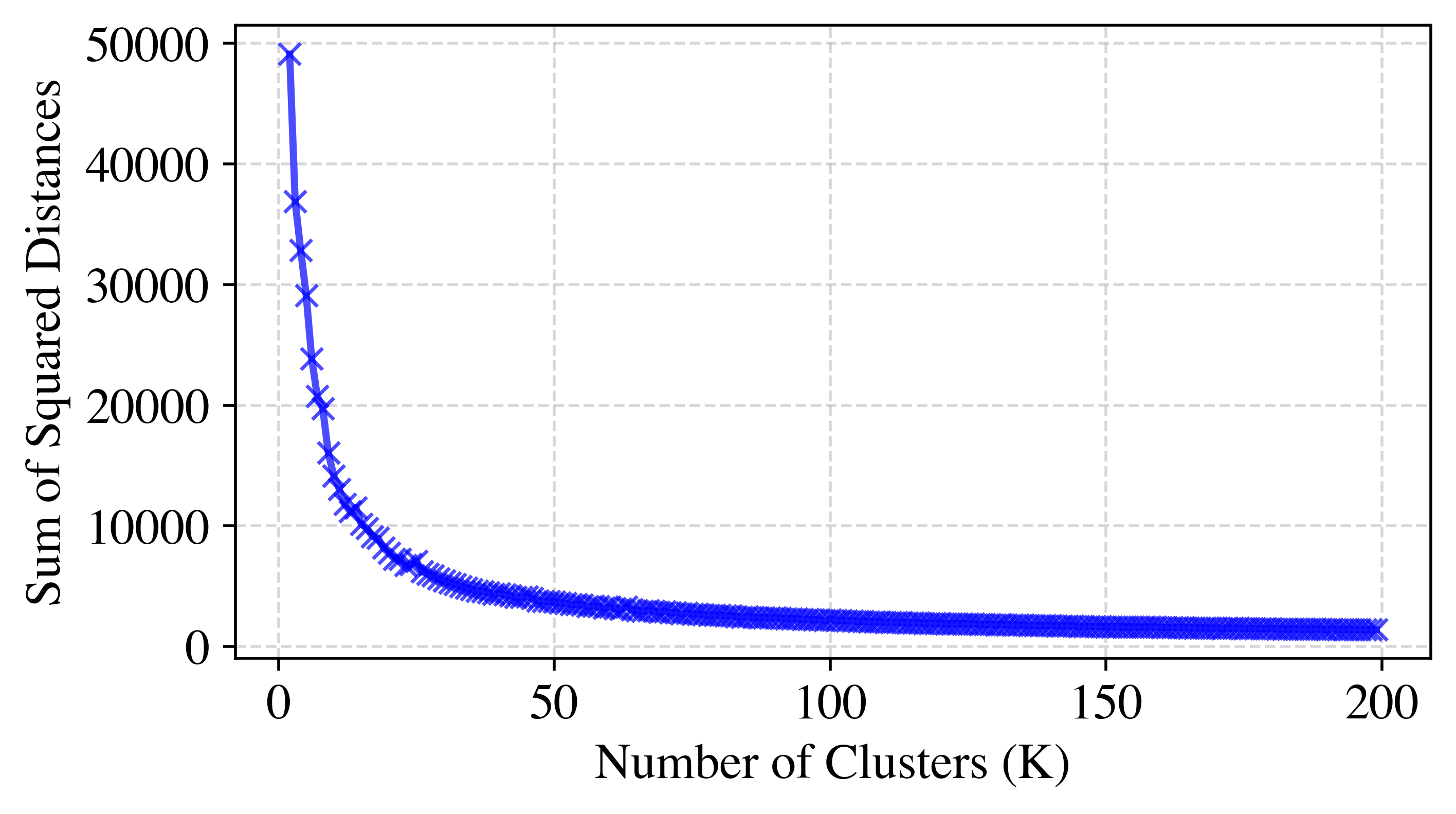}
    \vspace{-5pt}
    \caption{Result of the Elbow method to determine the optimal K clusters in BERTopic modeling.}
\label{fig:elbow}
\end{figure}

In this subsection, we present Figure~\ref{fig:elbow} to show how we decide the optimal number of clusters to conduct topic modeling.

\subsection{Geographical Distribution of Public Sentiment Across United States}
In this subsection, we present our visualization of the geographical distribution of public sentiments across the United States before and after the assassination attempt, as displayed in Figure~\ref{fig:gis}.

\subsection{Correlation Analysis between Demographic Factors and Average Public Sentiments}
\label{app:correlation}
As a complement, we conducted correlation analyses (as shown in Table~\ref{tab:corr}) to examine the relationship between socioeconomic and demographic factors and average public sentiments toward Donald Trump before and after the assassination attempt. States with higher Republican voter shares ($\mathit{Republican_R}$) displayed consistently more positive sentiment, with correlations of $0.448$ before and $0.538$ after the incident. Conversely, higher proportions of Democratic voters ($\mathit{Democrat_R}$) were associated with more negative sentiment, reflected in correlations of $-0.367$ and $-0.462$, respectively. Economic factors further highlight disparities in sentiment: areas with higher poverty rates ($\mathit{Household_Below_Poverty_R}$) and uninsured populations ($\mathit{No_Insurance_R}$) exhibited less negative sentiment, whereas states with higher median incomes ($-0.451$ before, $-0.568$ after) and bachelor’s degree attainment rates ($-0.495$ before, $-0.526$ after) expressed increasingly negative sentiment toward Trump.

While these correlation results provide valuable insights, it is important to note that they are based on data from only 50 states, which limits the strength and generalizability of the findings. The relatively small sample size means that some relationships may be influenced by regional outliers or other confounding factors not captured in the analysis. However, the consistency of key trends—such as the positive association between Republican vote share and sentiment, and the negative correlation with median income and bachelor’s degree attainment—partially validates the reliability of our sentiment analysis pipeline. These observed patterns align with broader sociopolitical divides documented in previous research, suggesting that our methodology is effective in capturing meaningful sentiment shifts within the dataset. This preliminary validation reinforces the pipeline’s utility for analyzing larger-scale social media data and identifying nuanced patterns in public discourse.

\begin{table}[htbp]
    \centering
    \small
    \caption{Correlation analysis of socioeconomic and demographic factors with sentiment before and after the assassination. Here, $\_R$ refers to rate.}
    \begin{tabular}{lrr}
    \hline
    \textbf{Variable} & \textbf{Before} & \textbf{After} \\
    \hline
    Total\_Population & -0.193 & -0.357 \\
    LUM\_Race & 0.173 & 0.018 \\
    Median\_income & -0.451 & -0.568 \\
    GINI & 0.303 & 0.199 \\
    Democrat\_R & -0.367 & -0.462 \\
    Republican\_R & 0.448 & 0.538 \\
    No\_Insurance\_R & 0.526 & 0.585 \\
    Household\_Below\_Poverty\_R & 0.560 & 0.595 \\
    HISPANIC\_LATINO\_R & -0.076 & -0.125 \\
    White\_R & -0.281 & -0.177 \\
    Black\_R & 0.347 & 0.236 \\
    Indian\_R & 0.076 & 0.230 \\
    Asian\_R & -0.337 & -0.447 \\
    Under\_18\_R & 0.252 & 0.381 \\
    Bt\_18\_44\_R & 0.033 & -0.102 \\
    Bt\_45\_64\_R & -0.246 & -0.237 \\
    Over\_65\_R & -0.093 & -0.040 \\
    Male\_R & 0.094 & 0.120 \\
    Bachelor\_R & -0.495 & -0.526 \\
    Population\_Density & -0.066 & -0.190 \\
    Unemployed\_R & 0.358 & 0.300 \\
    \hline
    \end{tabular}
    \centering
    \label{tab:corr}
\end{table}

\end{document}